\documentclass[aip,reprint,citeautoscript,floatfix]{revtex4-1}

\draft 

\usepackage{amsmath,amssymb,graphicx,hyperref}

\usepackage{xcolor}
\usepackage{hyperref}
\hypersetup{
colorlinks,
linkcolor={red!50!black},
citecolor={blue!80!black},
urlcolor={blue!80!black}
}

\providecommand{\ang}{\textup{\AA}}

\renewcommand{\eqref}[1]{Eq.~(\ref{#1})}
\providecommand{\Em}{E_{-}}
\providecommand{\Ep}{E_{+}}
\providecommand{\EN}{E_m} 
\renewcommand{\vec}[1]{\mathbf{#1}}

\begin{document}


\title{Lossless plasmons in highly mismatched alloys} 



\author{Hassan Allami}
\affiliation{Department of Physics, University of Ottawa, Ottawa, ON K1N 6N5, Canada}

\author{Jacob J. Krich}
\altaffiliation{School of Electrical Engineering and Computer Science, University of Ottawa, Ottawa, ON K1N 6N5, Canada}
\affiliation{Department of Physics, University of Ottawa, Ottawa, ON K1N 6N5, Canada}


\begin{abstract}
We explore the potential of highly mismatched alloys (HMAs) for realizing lossless plasmonics. Systems with a plasmon frequency at which there are no interband or intraband processes possible are called lossless, as there is no 2-particle loss channel for the plasmon. We find that the band splitting in HMAs with a conduction band anticrossing guarantees a lossless frequency window. When such a material is doped, producing plasmonic behavior, we study the conditions required for the plasmon frequency to fall in the lossless window, realizing lossless plasmons. Considering a generic class of HMAs with a conduction band anticrossing, we find universal contours in their parameter space within which lossless plasmons are possible for some doping range. Our analysis shows that HMAs with heavier effective masses are most promising for realizing a lossless plasmonic material.
\end{abstract}


\maketitle 


The field of plasmonics relies on surface plasmon polariton (SPP) modes, which can be excited on metal-dielectric interfaces \cite{Ritchie_plasma-theory,first-plasmon-exp}.
These SPP modes can enhance and concentrate electric fields at subwavelength scale \cite{maier2007plasmonics,SPP-optics,plasmonics-photonics-intro}, 
with applications in metamaterials \cite{roadmap_2016},
optoelectronics \cite{nonlinear-plasmonics,plasmonic-device-2013,spp-waveguide,spaser_2010,plasmonic-pv,plasmonic-sensor-review},
and photocatalysis \cite{plasmon-photcatalysis-review}. The applications range from the established surface-enhanced Raman scattering (SERS) technique \cite{SERS-original} to new proposals in quantum optics \cite{quantum-plasmonics} such as quantum teleportation \cite{spp-teleportation}.

In practice, many plasmonic applications are hindered by loss and decay of the plasmonic modes \cite{Boriskina_17,khurgin2012reflecting,Khurgin-loss-scaling}. 
Although some plasmonic applications such as SERS are still viable in the presence of losses \cite{khurgin2015deal},
and in cases such as photodetection and photocatalysis losses are beneficial \cite{Boriskina_17,plasmonic-sensor-review,plasmon-photcatalysis-review,thermo-plasmonic},
the loss problem is  one of the main challenges in the  field of plasmonics and metamaterials \cite{plasmonic-roadmap_18,roadmap_2016}.

Many approaches have been proposed to reduce and mitigate losses in plasmonic systems,
including improvements in fabrication \cite{surface-fab-loss-reduce},
employing optical gain \cite{Stockman_optical-gain},
spectrum modification \cite{fano-plasmonic},
and of course, searching for alternative plasmonic materials \cite{beyond-Au-Ag,searching-for-better,AlZnO-plasmonic,TiN-plasmonic,highly-dope-plasmonic,SC-plasmonic}.
Khurgin and Sun presented a strategy to find lossless plasmonic modes by considering the fundamental conditions creating  loss \cite{lossless_khurgin}.
Often the most important loss channel for SPPs is decay into electron-hole excitation. They argue that dissipating the energy of an electromagnetic mode in this way requires empty electronic states.
In a material with the appropriate electronic structure, such empty states may be absent for a range of energies, offering a lossless window of frequencies.
Hence they conclude that if the plasma frequency falls inside the lossless window, the primary decay mechanism will have been removed, producing an essentially lossless plasmonic material.
Khurgin and Sun proposed a few classes of materials that can potentially realize their conditions for lossless plasmons, and some of them have been investigated with promising results \cite{losslessAlO,layered_metal_2017,khurgin2017mitigating}. 

In this work, we propose highly mismatched alloys (HMA) as a candidate class for realizing mid- to far-IR lossless plasmons. This possibility was briefly mentioned but not elaborated in Ref.\ \onlinecite{rocksalt}.
We consider the one-particle and plasmonic structure of HMAs and find the alloying and doping requirements to achieve lossless plasmons. We describe universal contours in the parameter space of HMAs in which lossless plasmons are possible for some range of doping.
We show the alloy fractions and plasmon frequencies that occur in the lossless window for the ZnCdTeO system and suggest that HMAs with large effective masses are most likely to be able to realize the conditions needed for lossless plasmons.  
In the remainder of this work, we use the term ``lossless window'' in the sense of Ref.~\onlinecite{lossless_khurgin}.
			
\begin{figure}[ht]
	\includegraphics[width=\columnwidth]{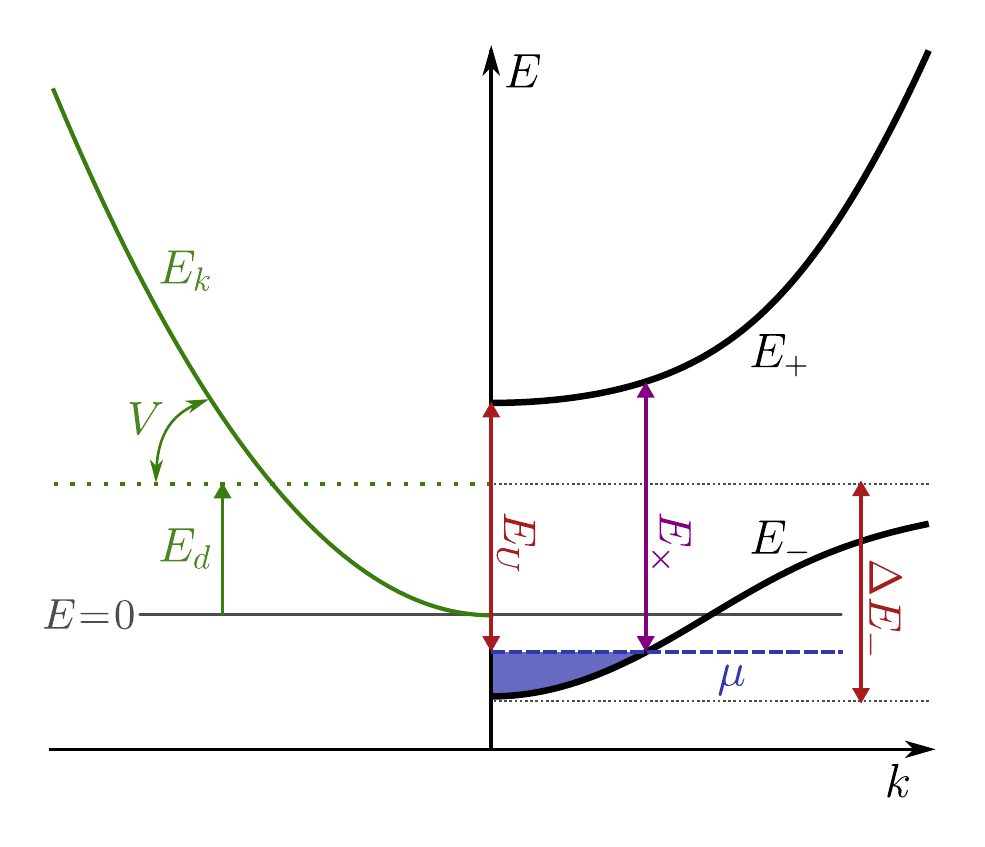}
	\caption{\label{fig:bands} (left) The conduction band of the host material $E_k$,
	and the localized state $E_d$, coupled through $V$. The energy reference is at the bottom of $E_k$.
	(right) The split bands of BAC model, $E_+$ and $E_-$.
	The interband loss starts for energies larger than $E_U$,
	and for the energies below $\Em$ bandwidth $\Delta E_-$,
	intraband dissipation is possible.
	Doping determines the chemical potential $\mu$.
	The interband transitions bound plasmon energy $\hbar\omega_p$ to
	$E_\times = \min_{k<k_F} \left(E^+_\vec{k} - E^-_\vec{k}\right)$.}
\end{figure}

HMAs are a class of semiconductor alloys where the alloying elements have very different electronegativity than that of the host.
Shan et al.\ described that localized states form around the mismatching elements and
proposed a band anticrossing (BAC) model, which successfully describes the energy spectrum of HMAs  \cite{bac-original}.
According to the BAC model, the localized level $E_d$ and the host conduction band (CB) with dispersion $E_\vec{k}$
hybridize  at each wavevector $\vec{k}$ independently with a single coupling factor $V$ (see Fig.~\ref{fig:bands}).
Two split bands emerge, with dispersion
\begin{equation}
	E_\pm = \frac{1}{2}\left(E_k + E_d \pm \sqrt{(E_k - E_d)^2 + 4V^2x}\right),
	\label{eq:Epm}
\end{equation}
where $x$ is the alloy fraction of the mismatching element. The two split bands $E_\pm$ are shown as solid black curves in Fig.~\ref{fig:bands}.
Here we consider a class of HMAs where the localized level anticrosses with a parabolic CB with $E_\vec{k} = \hbar^2k^2 / 2m$, the bottom of which is taken to be the zero energy level, as shown in Fig.~\ref{fig:bands}.
All such HMAs are described by three scalar parameters: $V\sqrt{x}$, $E_d$, and $m$.

\begin{figure}[ht]
	\includegraphics[width=\columnwidth]{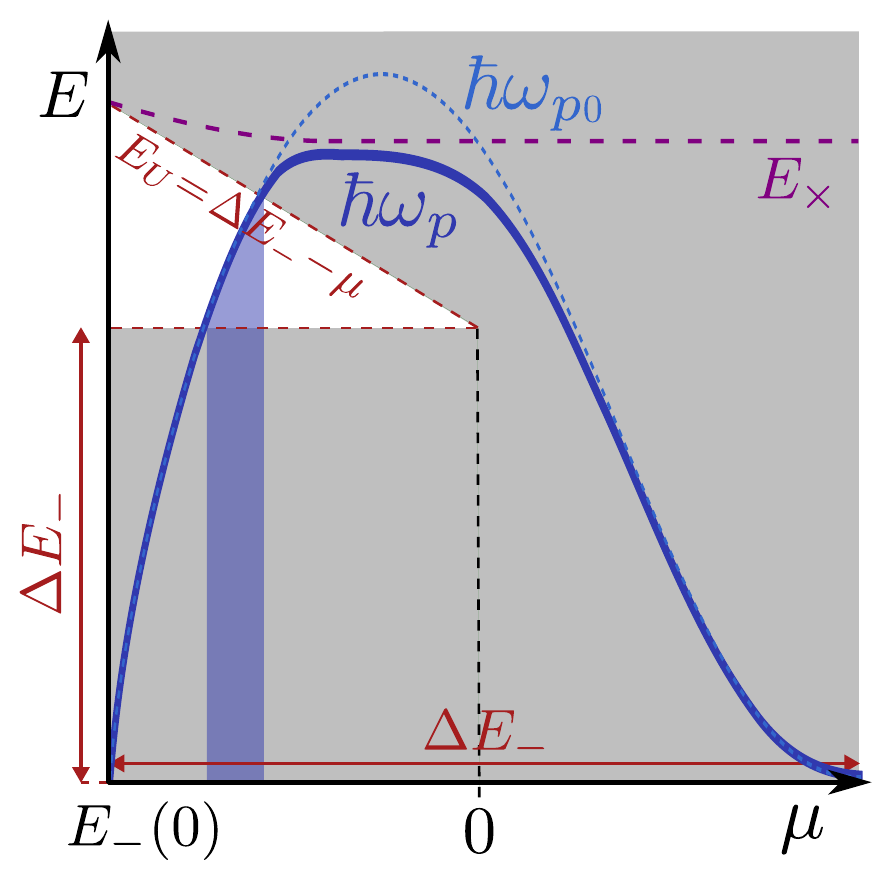}
	\caption{\label{fig:window} The lossless window of excitation energies (white area)
	and $\hbar\omega_p$ (dark blue line) as a function of chemical potential $\mu$.
	$\hbar\omega_p$ is bounded by $\hbar\omega_{p0}$ and $E_\times$.
	The blue shaded strip shows the range of $\mu$ where $\hbar\omega_p$ falls in the lossless window.
	}
\end{figure}

We now describe the range of excitation energies that can be lost to particle-hole excitations in a doped HMA. We consider that the excess electrons occupy the $\Em$ band, and we consider zero temperature, so the chemical potential $\mu$ lies somewhere in $\Em$, as in Fig.~\ref{fig:bands}.
The free electrons in $\Em$ produce plasmonic behavior \cite{HMA_plasmon} while at the same time providing two channels for dissipating energy.
First, an excitation of any amount of energy up to the bandwidth $\Delta\Em$ of the $\Em$ band can move an electron from a filled state to an empty state within the $\Em$ band. The excited electron and hole can then dissipate their energy into a set of excitations with infinitesimal energy by moving electrons near the Fermi surface to nearby empty states. 
Figure \ref{fig:window} shows the range of excitation energies and $\mu$ where dissipation into particle-hole excitations can occur; the grey area under the horizontal dashed red line represents this intraband lossy region.
Second, an electron can be excited to the $\Ep$ band.
The minimum energy for such interband transitions is $E_U\equiv \Ep(k=0) - \mu$, which is labeled in Figs.~\ref{fig:bands} and \ref{fig:window}.
Excitations with energy higher than $E_U$ can then be lost through a combination of inter- and intra-band transitions.
The grey area above the slanted dashed line in Fig.~\ref{fig:window} represents this second lossy region. 

Excitations with energy between $\Delta\Em$ and $E_U$ are lossless, i.e., have no single-particle decay channels, if $E_U > \Delta\Em$.
The white area in Fig.~\ref{fig:window} shows this lossless window,
which linearly shrinks with increasing $\mu$.
Measuring energies from the bottom of $E_\vec{k}$, it turns out that $\Ep(k=0)=\Delta\Em$ , so $E_U = \Delta\Em - \mu$.
So there is always a lossless window if $\mu < 0$.
Moreover, since the minimum of $\Em$ is always negative,
for any HMA there is always a doping level
below which there exists a lossless window.

The next step is to check when the plasmon energy $\hbar\omega_p$ falls in the lossless window.
$\hbar\omega_p$ depends on the carrier concentration and hence on $\mu$, which ranges between the bottom and the top of $\Em$ at zero temperature.
As shown by the blue curve in Fig.~\ref{fig:window}, $\hbar\omega_p$ initially rises with $\mu$ as free carriers enter the $\Em$ band and falls back to zero when the band is full \cite{HMA_plasmon}.
Depending on the HMA parameters, $\hbar\omega_p$ can fall in the lossless window for some range of doping, as in the case in the figure.

In previous work \cite{HMA_plasmon}, we found that $\omega_p$ for this class of HMAs obeys
\begin{equation}
	\omega_p^2 = \omega_{p0}^2 \left[1+ \left(\frac{mc\ell}{\hbar}\right)^2 I_\times(\omega_p)\right]^{-1},
	\label{eq:wp}
\end{equation}
in which $\omega_{p0}$ is the plasma frequency in the absence of interband transitions,
and the second term in the bracket represents the effect of interband transitions,
where $\ell$ is a length scale that determines the strength of the transitions, 
which needs to be determined for each HMA. We provide more details in Section A of the supplementary material.
Ref.\ \onlinecite{HMA_plasmon} derived a closed algebraic form for $\omega_{p0}$ and an integral for $I_\times(\omega)$,
which is always positive and diverges as
$\hbar\omega$ approaches $E_\times \equiv \min_{k<k_F} \left(E^+_\vec{k} - E^-_\vec{k}\right)$, where $k_F$ is the Fermi momentum. 
Therefore, $\hbar\omega_p$ is bounded by $\hbar\omega_{p0}$ and $E_\times$, as Fig.~\ref{fig:window} shows.
Note that since $E_\times$ is the minimum distance between $\Ep$ and the filled part of $\Em$ at the same $\vec{k}$, it is always larger than $E_U$, which does not have the same $\vec{k}$ restriction (see Figs.~\ref{fig:bands} and \ref{fig:window}).

\begin{figure}[ht]
	\includegraphics[width=\columnwidth]{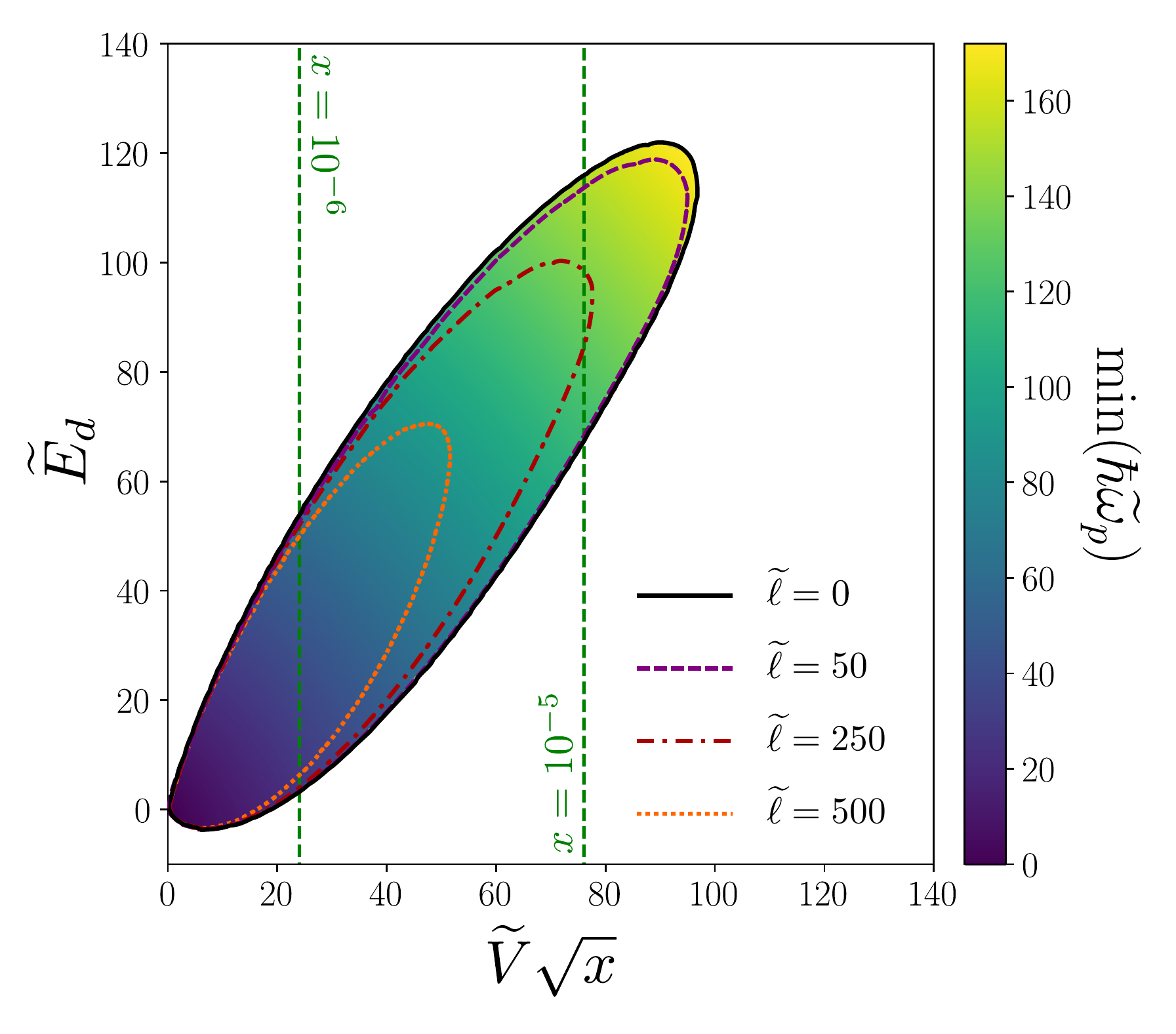}
	\caption{\label{fig:region} Universal contours showing regions with lossless plasmons 
	for several $\widetilde{\ell} \equiv mc \ell / \hbar$, with energies all normalized by $\EN\equiv 2mc^2\times10^{-9}$.
	The color scale shows the smallest normalized plasmon energy in the lossless window, which is always equal to $\Delta\Em/\EN$, independent of $\widetilde{\ell}$.
	The green almost-vertical dashed lines show the locus of Zn$_{1-y}$Cd$_y$Te$_{1-x}$O$_x$,
	for $x = 10^{-6}$ and $x = 10^{-5}$, as $y$ changes approximately between 0.27 and 0.29. 
	}
\end{figure}

If we normalize all energies to $2mc^2$,
then we can accommodate this entire  class of HMAs in a 2D plane spanned by $V\sqrt{x}/2mc^2$ and $E_d/2mc^2$. But since $2mc^2$ for typical semiconductors is of the order of MeV, while meV is a more relevant unit for typical $V\sqrt{x}$ and $E_d$, we normalize energies using $\EN\equiv2mc^2\times 10^{-9}$ and define $\widetilde{V}\equiv V/\EN$, $\widetilde{E}_d\equiv E_d/\EN$. For the bare electron mass $m_e$,  $\EN\approx 1$~meV, so
for any other effective mass $m$ one can scale all dimensionless energies by $m/m_e$ to find the approximate value in meV.

We can determine for any point in the $(\widetilde{V}\sqrt{x}, \widetilde{E}_d)$ plane whether there is a doping range in which $\hbar\omega_p$ falls in the lossless window, realizing lossless plasmons.
Then, with fixed $\widetilde{\ell} \equiv \ell mc/\hbar$, where $\hbar/mc$ is the reduced Compton wavelength, 
there is a universal contour in the $(\widetilde{V}\sqrt{x}, \widetilde{E}_d)$ plane within which such HMAs fall.
Fig.~\ref{fig:region} shows these universal contours of lossless plasmonic HMAs for a few values of $\widetilde{\ell}$. The solid black contour 
shows the $\ell=0$ case without interband transitions, where $\omega_p=\omega_{p0}$.
The contours do not vary significantly from the $\widetilde{\ell}=0$ case until $\widetilde{\ell}\gtrsim100$ 
since  $I_\times(\omega)$ in \eqref{eq:wp} is significant only for $\hbar\omega_p$ near $E_\times$. 
But since in the lossless window  $\hbar\omega_p<E_U<E_\times$, 
the interband processes can move $\hbar\omega_p$ away from $\hbar\omega_{p0}$ only when $\widetilde{\ell}$ is large.
For estimated typical values of $\ell=1-10$~$\ang$, $\widetilde\ell \approx (250-2500) m/m_e$. 
Using $\omega_{p0}$ in place of $\omega_p$, as in the $\widetilde{\ell}=0$ case, allows us to derive an analytic expression for the lossless contour, which is presented in  Section B of the supplementary material.

The color scale of Fig.~\ref{fig:region} shows the smallest $\hbar\widetilde\omega_p$ in the lossless window.
As Fig.~\ref{fig:window} shows, the minimum value of $\hbar\omega_p$ in the lossless window is always $\Delta\Em$, which does not depend on $\ell$.
Therefore, $\min(\hbar\omega_p/\EN)$
is the same for lossless contours belonging to different $\widetilde{\ell}$.
Although the largest $\hbar\widetilde\omega_p$ in the lossless window is somewhat different for each $\widetilde{\ell}$, the typical difference between $\max(\hbar\omega_p)$ and $\min(\hbar\omega_p)$ is of the order of a few $\EN$ for all cases.

Since $\EN\propto m$, Fig.~\ref{fig:region} shows that HMAs with heavier $m$ can achieve lossless plasmons for a wider range of $V\sqrt{x}$ and $E_d$.
Since the typical values of $V$ are on the order of eV, increasing the range of $V\sqrt{x}$ is particularly crucial because otherwise the required $x$ may be too small to realistically show alloying effects.

To illustrate this point, consider the quaternary HMA, Zn$_{1-y}$Cd$_y$Te$_{1-x}$O$_x$,
in which oxygen is the mismatching element in ZnCdTe.
Table~\ref{tab:ZnCdTeO} shows the range of BAC parameters for this quaternary, determined from studies of alloy-dependent band gap.\cite{Tanaka_2016,adachi_mass}
Doping of the $\Em$ band with chlorine has been demonstrated \cite{cl-dope-ZnTeO-19}, and this system can realize different values of $E_d$ by tuning the cadmium content $y$.
The $E_d$ range covered in Fig.~\ref{fig:region} corresponds to $y$ between 0.27 and 0.29. 
The two nearly vertical dashed green lines in Fig.~\ref{fig:region} show the locus of ZnCdTeO in the HMA parameter space, as $y$ varies for two fixed values of oxygen fraction, $x = 10^{-6}$ and $x = 10^{-5}$, with details of the assumed bowing parameters described in  Section~C of the supplementary material.
This range of $y$, which produces $E_d$ near zero, allows overlap of $\omega_p$ with the lossless region when $x$ is sufficiently small.
As the smallest $x$ for which the BAC model applies is unexplored, it is not clear whether these particular parameter ranges are experimentally realizable.
ZnCdTe has $m\approx0.1m_e$, so in the lossless window we find 
$\hbar\omega_p$ from 3 to 6~meV for $x=10^{-6}$ and 13 to 16~meV for $x=10^{-5}$.
For these small $\hbar\omega_p$, phonons could directly couple to the plasmons, providing a new channel for dissipation and breaking the lossless condition. A material with a heavier effective mass would allow both for larger $x$ and lossless plasmons with higher $\hbar\omega_p$.

Observing these low plasma frequencies may require low temperatures, as \eqref{eq:wp} is derived at $T=0$. At higher temperatures, the occupation fraction of states in the $\Em$ band must be taken into account.
Nonetheless, the suppression of electronic decay channels in this lossless window should still be visible in experiments.

\begin{table}[t]
	\caption{
	The range of BAC parameters for Zn$_{1-y}$Cd$_y$Te$_{1-x}$O$_x$.\cite{Tanaka_2016,adachi_mass}
	}
	\label{tab:ZnCdTeO}
	\begin{ruledtabular}
	\begin{tabular}{lcc}
		Parameters & $y = 0$ & $y = 1$\\
		\hline
		$E_d$ [eV] & -0.27 & 0.38\\
		$V$ [eV] & 2.8 & 2.2\\
		$m$ [$m_e$] & 0.117 & 0.09\\
	\end{tabular}
	\end{ruledtabular}
\end{table}

Although overall HMAs with heavier $m$ are more promising in realizing lossless plasmons, not everything favors them.
Interband transitions shrink the lossless contour, as shown in Fig.\ \ref{fig:region}, and
$\widetilde{\ell}$ increases linearly with $m$. 
Lighter $m$ can allow smaller $\widetilde\ell$ and a larger lossless contour. 

The two most explored classes of conduction band HMAs are III-V nitrides and II-VI oxides \cite{HMA-special-topic}.
Given the low effective masses of the conduction bands in these materials, the lossless plasmonic window will occur for THz-frequency $\omega_p$ with extremely light alloying.
As new HMAs with larger effective masses in their host bands are found, they will provide further opportunities to realize lossless plasmonic materials. 
HMAs with valence band anticrossings \cite{InAlBiAs-HMA,GaInAsBi-HMA} could be good candidates, as they typically have heavier effective masses, though the theory of their plasmon frequencies has not yet been worked out.
HMAs present exciting potential for realizing a lossless plasmonic medium and are worth more theoretical and experimental investigations.
\\


%

See the supplementary material for the details of the plasma frequency equation,
the analytic expression of the lossless contour for the case of $\ell = 0$,
and the locus of ZnCdTeO in the HMA parameter space.
\\

We acknowledge funding from the NSERC CREATE TOP-SET program, Award Number 497981.
\section*{author declarations}
\subsection*{Conflict of Interest}
The authors have no conflicts to disclose.
\section*{data availability}
The data that support the findings of this study are openly available at \url{https://github.com/hassan-allami/Lossless-HMAs} \cite{codes_repo}.

\appendix
\begin{widetext}

\section{Deriving Eq.\ (2) for the plasma frequency \label{sec:wp}}
\eqref{eq:wp} of the main text is the rewriting of Eq.~(20) of Ref.~\onlinecite{HMA_plasmon}.
In Ref.~\onlinecite{HMA_plasmon}, $\omega_{p0}$ is defined in Eq.~(18) in terms of $E_{k_F} = \hbar^2k_F^2/2m$.
By solving $\Em |_{k = {k_F}} = \mu$ for $E_{k_F}$, one can find $E_{k_F} = \mu + V^2x / (E_d - \mu)$. 
This form allows rewriting Eq.\ (18) of Ref.\ \onlinecite{HMA_plasmon} to give $\hbar\omega_{p0}(\mu)$ as
\begin{equation}
	\hbar\omega_{p0} = \sqrt{\frac{8\alpha}{3\pi}}(2mc^2)^{1/4}
	\left(\frac{V^2x}{E_d - \mu} + \mu \right)^{3/4}
	\left(\frac{(E_d - \mu)^2}{(E_d - \mu)^2 + V^2 x}	\right)^{3/2},
	\label{eq:wp0}
\end{equation}
where $\alpha = e^2 / \hbar c$ is the fine structure constant.

The dimensionless $I_\times$ can be expressed as 
\begin{equation}
	I_\times(\widetilde{\omega}) = \frac{4\alpha \widetilde{V}^2 x}{\pi}\int_0^{\widetilde{k}_F}
	\frac{\widetilde{k}^2 d\widetilde{k}}
	{(\widetilde{E}_+ - \widetilde{E}_-)[(\widetilde{E}_+ - \widetilde{E}_-)^2 - (\hbar\widetilde{\omega})^2]},
	\label{eq:Ix}
\end{equation}
where all energies are in units of $\EN\equiv 2mc^2\times10^{-9}$ and all wavevectors are in units of $k_m\equiv mc/\hbar\times10^{-3}$. 
This form shows that $I_\times$ is proportional to $\alpha$, so it is generally small except near where it diverges. 
It diverges at the smallest $\omega$ where $\hbar\omega = \min(\Ep - \Em)$ for some $k$ in the integration domain,
which motivates the definition of $E_\times = \min_{k<k_F}(\Ep - \Em)$.

\section{Finding the lossless contour without interband transitions \label{sec:contour}}
To find the lossless contour for the case of $\ell = 0$, we need to find the set of HMA parameters for which $\hbar\omega_{p0}$ falls in the lossless window for any $\mu$.
As Fig.~\ref{fig:window} shows, $\hbar\omega_{p0}(\mu)$ has a maximum. We start by locating this maximum point $\mu_{\max}$.
It is more convenient to find the maximum point for $(\hbar\omega_{p0})^4$, which is also $\mu_{\max}$. We define $\lambda = \widetilde{E}_d/\widetilde{V}\sqrt{x}$ and $\nu = \widetilde{\mu}/\widetilde{V}\sqrt{x}$. Then using \eqref{eq:wp0} to solve $\partial(\hbar\widetilde{\omega}_{p0})^4/\partial\nu =0$ gives, after dividing out constants,
\begin{equation}
	\frac{(\lambda-\nu)^8(\nu^2 - \lambda\nu -1)^2
	[\nu^4 - 4\lambda\nu^3 + 6(\lambda^2 + 1)\nu^2-4(\lambda^3 + 2\lambda)\nu +(\lambda^2 + 3)(\lambda^2 - 1)]}
	{\left[(\lambda-\nu)^2 + 1\right]^7}
	= 0.
	\label{eq:derivative}
\end{equation}
Since the largest $\mu$ in the lossless window is $\mu=0$ (see Fig.~\ref{fig:window}), we must find whether $\mu_\text{max}$ is negative. 

When $E_d > 0$, note from \eqref{eq:derivative} that when $\nu = \mu = 0$,
which is the right edge of the lossless window,
the sign of \eqref{eq:derivative} is determined by the sign of $\lambda - 1$.
So $\mu_{\max} \geq 0$ if $\lambda \geq 1$.
In this case $\max_{\mu<0}(\hbar\omega_{p0}) = \hbar\omega_{p0}|_{\mu=0}$.
Note that the lossless window is open only for $\mu < 0$ and requires $\hbar\omega_p>\Delta\Em$, 
so in this case $\hbar\omega_{p0}$ can fall inside the window if $\hbar\omega_{p0}|_{\mu=0} > \Delta\Em$.
Therefore, if $E_d \geq V\sqrt{x}$, the lossless plasmonic region in the HMA parameter space is determined by
\begin{equation}
	\sqrt{\frac{8\alpha}{3\pi}}\left(\frac{2mc^2}{\EN}\right)^{1/4}
	\left(\frac{\widetilde{V}^2x}{\widetilde{E}_d}\right)^{3/4}
	\frac{\widetilde{E}_d^3}{(\widetilde{E}_d^2 + \widetilde{V}^2x)^{3/2}} >
	\frac{\widetilde{E}_d + \sqrt{\widetilde{E}_d^2 + 4\widetilde{V}^2x}}{2},
	\label{eq:lambda>1}
\end{equation}
where we used \eqref{eq:wp0} for $\hbar\omega_{p0}$, and $\Delta\Em = \frac{1}{2}(E_d + \sqrt{E_d^2 + 4V^2x})$, which can be derived from \eqref{eq:Epm}.

When $E_d <V\sqrt{x}$ and hence $\mu_{\max} < 0$,  then the lossless window is open at $\mu_\text{max}$, 
so $\hbar\omega_{p0}$ falls in the lossless window when
$\max(\hbar\omega_{p0})>\Delta\Em$. For $E_d < 0$,
we always have $\mu_{\max} < 0$, because $\mu_\text{max}$ must fall within the $\Em$ band and $E_d<0$ is the top of the $\Em$ band.

To find $\mu_{\max}$, notice that $(\lambda-\nu)^8(\nu^2 - \lambda\nu -1)^2$ is zero only when $\mu$ is at the bottom or top of the $\Em$ band, where $\omega_{p0}~=~0$. Then Eq.\ \ref{eq:derivative} says that $\mu_\text{max}$ is determined by 
the real roots of $[\nu^4 - 4\lambda\nu^3 + 6(\lambda^2 + 1)\nu^2-4(\lambda^3 + 2\lambda)\nu +(\lambda^2 + 3)(\lambda^2 - 1)]$.
This term always has two real roots, and only one of them corresponds to a $\mu$ inside the $\Em$ band.
Picking the relevant root and using $\widetilde{\mu} = \nu\widetilde{V}\sqrt{x}$, we obtain
\begin{align}
	 \label{eq:u_max}
	& \widetilde{\mu}_{\max} = \widetilde{E}_d \left(1 - \sqrt{\frac{\widetilde{V}^2 x A}{\widetilde{E}_d^2}}\right)
	- \sqrt{\widetilde{V}^2x B},  \\
	&A = \left(1 + \frac{\widetilde{E}_d^2}{4\widetilde{V}^2x} \right)^{1/3} - 1, \nonumber\\
	&B = \sqrt{\frac{\widetilde{E}_d^2}{\widetilde{V}^2 x A}} - A - 3. \nonumber
\end{align}

Using \eqref{eq:u_max} for $\mu_{\max}$ in \eqref{eq:wp0}, we find $\max(\hbar\widetilde{\omega}_{p0})$,
which gives an explicit expression for the lossless plasmonic region when $\widetilde{E}_d < \widetilde{V}\sqrt{x}$,
\begin{equation}
	\sqrt{\frac{8\alpha}{3\pi}}\left(\frac{2mc^2}{\EN}\right)^{1/4}
	\left(\frac{\widetilde{V}^2x}{\widetilde{E}_d - \widetilde{\mu}_{\max}} + \widetilde{\mu}_{\max}\right)^{3/4}
	\left(\frac{(\widetilde{E}_d - \widetilde{\mu}_{\max})^2}
	{(\widetilde{E}_d - \widetilde{\mu}_{\max})^2 + \widetilde{V}^2x}\right)^{3/2} >
	\frac{\widetilde{E}_d + \sqrt{\widetilde{E}_d^2 + 4\widetilde{V}^2x}}{2}.
	\label{eq:lambda<1}
\end{equation}

The combination of \eqref{eq:lambda>1} and \eqref{eq:lambda<1}  defines the lossless region in the HMA parameter space for the case of $\ell=0$, shown by the solid black contour in Fig.~\ref{fig:region} of the text.

\section{BAC parameters of Zn$_{1-y}$Cd$_y$Te$_{1-x}$O$_x$ \label{sec:contour}}
Table~\ref{tab:ZnCdTeO} shows the BAC parameters of Zn$_{1-y}$Cd$_y$Te$_{1-x}$O$_x$ for $y=0$ and $y=1$ extracted primarily from optical measurements \cite{Welna_2015,Yu_2004,Seong_1998,Tanaka_2016,adachi_mass}. 
We include bowing for $E_d$ using
$E_d(y) = (1-y)E_d\vert_{y = 0} + yE_d\vert_{y = 1}+ y(1-y)C$,
with $C~=~0.46$~eV \cite{SAMANTA_1995,Tanaka_2016},
while for $V$ and $m$, we used linear interpolation between the values shown in Table~I.

The two dashed green lines in Fig.~\ref{fig:region} show the locus of ZnCdTeO for two different fractions of oxygen as $y$ changes.
The lines are not perfectly vertical, but they are nearly so because $V\sqrt{x}$ varies much less than $E_d$ for the same range of $y$. Also, both $V$ and $m$ decrease with a similar rate as $y$ increases, leaving $V/\EN\propto V/m$ nearly unchanged.

\end{widetext}

\section*{references}
\bibliography{references}

\end{document}